\documentstyle[aps]{revtex}
\input psfig.sty 
 
\begin{document}
\draft

\twocolumn[\hsize\textwidth\columnwidth\hsize\csname @twocolumnfalse\endcsname
\bibliographystyle{plain}
\title{ 
Persistent currents in a Moebius ladder: 
A test of interchain coherence of interacting electrons}
\vskip0.5truecm 
\author {Fr\'ed\'eric Mila$^{a}$, Charles Stafford$^{b}$ and 
Sylvain Capponi$^{a}$} 
\vskip0.5truecm
\address{
      $(a)$ Laboratoire de Physique Quantique, Universit\'e Paul Sabatier,
      31062 Toulouse (France)\\
      $(b)$ Institut de Physique Th\'eorique,
      Universit\'e de Fribourg, 1700 Fribourg (Switzerland)
      }
\maketitle

\begin{abstract}
Persistent currents in a Moebius ladder are shown to be very sensitive to the
effects of intrachain interactions on the hopping of electrons between chains.
Their periodicity as a function of flux is doubled 
for strong enough repulsive interactions
 because electrons cannot hop coherently between the
chains and have to travel along the full edge of the Moebius ladder, thus
encircling the flux twice. 
Mesoscopic
devices that should enable one to observe these effects are proposed.
\end{abstract}

\vskip.1truein 


\vskip2pc]

\narrowtext
The problem of transport in very anisotropic systems has become one of the
central issues in the field of strongly-correlated systems. Roughly speaking, it
can be stated as follows: Consider an electronic
system with transfer integrals much
smaller in one direction than in the other(s). Is it possible, and
under which conditions, that transport is coherent 
in the highly conducting direction(s) and incoherent
in the other direction at temperatures much smaller than the
smallest transfer integrals? This question has recently been raised in several 
contexts, the most prominent examples being the high-T$_c$ superconductors for
quasi-2D systems\cite{clarke,chakravarty}, and the organic superconductors, 
the Bechgaard salts, for
quasi-1D systems\cite{strong,danner}. This property is clearly inconsistent 
with Fermi Liquid
theory, and the best candidates to describe such a behaviour are models of
strongly-correlated electrons. Metallic behaviour is characterized by
the presence of a Drude peak in the optical conductivity at zero temperature,
and what one is looking for is a model for which
there is no Drude peak in the conductivity in one direction, while there is one
in the other(s). The optical conductivity is not a simple object to calculate 
though, and a number of authors concentrate on the Green's functions and look
for models with no pole in the Green's function in one direction and one in the
other(s). Both properties are very likely related, although this is not totally
clear due to vertex corrections.

So far, most results have been obtained for quasi-1D systems, i.e. systems of
coupled Luttinger liquids. Renormalization Group (RG)
arguments\cite{bourbonnais} suggest 
that there
will be a pole in the transverse Green's function as long as the Luttinger 
liquid exponent $\alpha$ is smaller than 1, while another approach drawing an
analogy to the problem of coherence in a two-level system coupled to a bath
suggests that the pole in the transverse Green's functions might disappear much
earlier\cite{proceedings}. 
Evidence in favor of the latter picture 
based on the analysis of the angular dependence of the magnetoresistance of the
Bechgaard salt (TMTSF)$_2$PF$_6$ has been presented \cite{strong,danner}. 
However, the issue remains controversial, and 
more direct evidence of the effect of
correlations on the coherence of motion perpendicular to the chains 
would be highly desirable.
In the case of two chains, coherence shows up as a splitting of the 
bonding and antibonding bands, which could in principle be measured in
angular resolved photoemission experiments. However, considering 
the difficulty in interpreting
photoemission experiments on low dimensional systems, and the need to vary
parameters like $\alpha$ or the perpendicular hopping, this prospect remains
remote to say the least.

\begin{figure}[hp]
\centerline{\psfig{figure=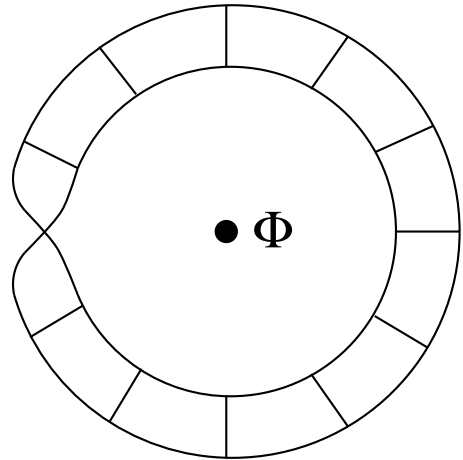,width=6.0cm,angle=0}}
\vspace{0.5cm}
\caption{Schematic diagram of a Moebius ladder pierced by a magnetic
flux $(\hbar c/e)\Phi$.}
\label{fig1}
\end{figure}

In this paper, we propose a new approach to study the problem of interchain 
coherence based on the analysis of persistent currents in a Moebius ladder
(see Fig.\ 1). 
If electrons can hop coherently from one edge to the other,
the groundstate energy and the persistent current will 
be  periodic functions of the flux with the usual
period $\phi_0=hc/e$. 
However, if the perpendicular hopping integral $t_\perp$ is
switched off, electrons will have to go twice around the flux to reach the same
site, and the period will be $\phi_0/2$. Now, for an interacting system with
$t_\perp\ne 0$, the period is expected to be $\phi_0$ as long as coherent motion
between the chains is possible, and to become $\phi_0/2$ 
when interactions are strong
enough to prevent any coherent motion between the chains. Thus, measuring the
periodicity 
of the persistent current should provide direct information on the
coherence of interchain hopping.

\begin{figure}[hp]
\centerline{\psfig{figure=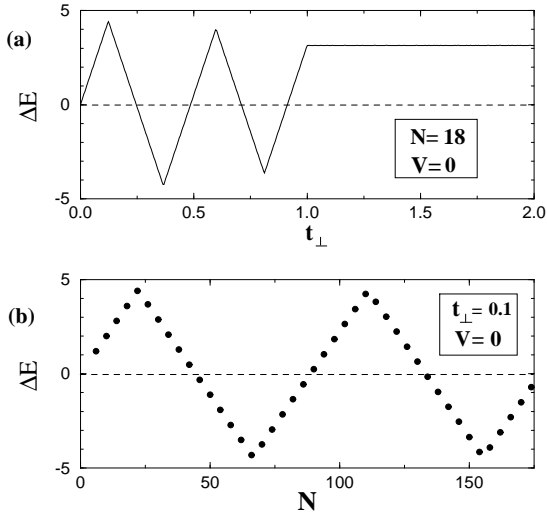,width=8.0cm,angle=0}}
\vspace{0.5cm}
\caption{$\Delta E$ for a non-interacting,
quarter filled system: a) as a function of $t_\perp$; b) as a function of the
number of rungs $N$.}
\label{fig2}
\end{figure}

This picture turns out to be essentially true, although the analysis is not as
straightforward as one might hope due to rather interesting finite-size
effects. Let us start with a careful analysis of the non-interacting case.
For simplicity, we consider spinless fermions throughout this paper. The spin is
not expected to play any significant role for this property, a point of view
confirmed by preliminary results we have obtained for fermions with spin.
Spinless fermions on a Moebius ladder with $N$ rungs 
pierced by a magnetic flux $(\hbar c/e)\Phi$ 
can be described by a one-dimensional periodic Hamiltonian 
\begin{eqnarray}
H & = & -t\sum_{i=1}^{2N} 
( e^{i\Phi/N} c^\dagger_i c_{i+1}^{\vphantom{\dagger}} 
+ h.c.)\nonumber \\
 & - & {1\over 2} t_\perp \sum_{i=1}^{2N} 
 ( c^\dagger_i c_{i+N}^{\vphantom{\dagger}}
  + h.c.) 
\end{eqnarray}
with the usual convention $c_{i+2N}=c_{i}$. This
Hamiltonian 
is readily diagonalized by a Fourier transform, and the dispersion
reads 
\begin{equation}
\epsilon_k=-2t\cos (ka+\Phi/N) -t_\perp \cos(Nka)
\nonumber
\end{equation}
with $k=p (2\pi/2Na)$, $p$ integer. The essential ladder structure of the
Moebius ladder is contained in this expression because 
$\cos(Nka)=\cos(p\pi)=+1$ if $p$
is even and $-1$ if $p$ is odd. The system thus consists of  bonding and 
antibonding bands with the usual dispersions 
$\epsilon_k=-2t\cos (ka+\Phi/N) \pm 
t_\perp$. The difference with a standard ladder is that the wavevectors 
$k=p (2\pi/2Na)$ are restricted to even and odd values of $p$ for the 
bonding and antibonding bands respectively.

Let us consider the periodicity of the groundstate energy of such a system as a
function of $\Phi$. If $t_\perp$ is large enough, all the fermions
are in the bonding band, and the total energy reads $E(\Phi)=\sum_p  
[-2t \cos (p\pi/N + \Phi/N) -t_\perp]$, where the sum over $p$ is restricted to
{\it even} integers chosen 
to give the lowest energy for a given $\Phi$. This function is clearly periodic
in $\Phi$ 
with period $2\pi$. If $t_\perp=0$, the bonding and antibonding bands form a
single band, and the total energy reads $E(\Phi)=-2t\sum_p  
\cos (p\pi/N + \Phi/N)$, where the integers $p$ can now be both
even and odd, and the periodicity of this function is $\pi$. 
For intermediate values of $t_\perp$, the periodicity is $2\pi$ except for 
specific values of $t_\perp$ where it is $\pi$. The number of such points scales
with the number of particles in the system. 
To understand this, let us consider the quantity
\begin{equation}
\Delta E=(-1)^{n-1 \over 2} N [ E(\Phi=\pi)-E(\Phi=0) ]
\end{equation}
where $n$ is the number of particles. In the following, $n$ is restricted to odd
values to have a non degenerate groundstate which makes the analysis slightly
simpler, although the results are essentially equivalent for an even number of
particles, and the factor $(-1)^{n-1 \over 2}$ has beeen included to insure 
that $\Delta E$ is always
positive in the limit $t_\perp \rightarrow 0$. Besides, as for 
the curvature at $\Phi=0$, which gives the Drude weight, one has to 
multiply by $N$ 
for 1D systems to get non-vanishing
results in the thermodynamic limit. 
A typical example of the behaviour of $\Delta E$ with $t_\perp$ is
depicted in Fig. 2a. It is clear from the dispersion of Eq. (2) that both 
$E(\Phi=\pi)$ 
and $E(\Phi=0)$, and hence $\Delta E$, are piece-wise linear functions 
of $t_\perp$. The slope of $\Delta E$ changes each time a pair of 
particles goes from the antibonding band 
to the
bonding band, which we know has to occur because the antibonding band is
certainly empty for large enough $t_\perp$. Now these transitions do not occur
for the same values of $t_\perp$ for $\Phi=\pi$ and $\Phi=0$. 
As a result, the slope of $\Delta E$ alternates
between $2N$ and $-2N$ until the antibonding band is empty, in which
case it is of course 0. Between changes of slopes, $\Delta E$ vanishes once,
which corresponds to points where the periodicity is $\pi$, like for
$t_\perp=0$. Since the particles change bands
in pairs, the number of such points is
essentially half the number of particles (more precisely $2n+1$ including 
$t_\perp=0$ for $4n+1$ particles, and similar formulae for other fillings).
The same effect shows up in 
the behaviour of $\Delta E$ with $N$:
For intermediate values of $t_\perp$, $\Delta E$ oscillates as a function of 
size
with extrema each time the difference in the particle number between the bonding
and the antibonding bands increases by 2 (see Fig. 2b). The points lie on a
piece-wise linear curve with a slope alternating
between $2t_\perp$ and $-2t_\perp$. As a consequence, the periodicity of the 
groundstate 
energy for a given $t_\perp$ and for non interacting fermions is not a well
defined quantity in the thermodynamic limit.

\begin{figure}[hp]
\centerline{\psfig{figure=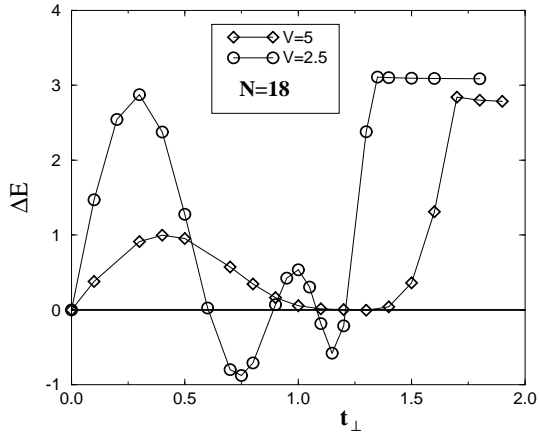,width=8.0cm,angle=0}}
\vspace{0.cm}
\caption{$\Delta E$ as a function of $t_\perp$ for $V=2.5$
($\alpha=0.36$) and $V=5$ ($\alpha=1$) for a quarter-filled system with 18
rungs. All energies are in units of $t$.}
\label{fig3}
\end{figure}

Let us now consider the effect of intrachain interactions on the periodicity of
the groundstate energy. To describe systems with large values of the Luttinger
liquid exponent $\alpha$, we consider an interaction term of the form
\begin{equation}
H_{\rm int}=\sum_i \left ( V  n_i n_{i+1} 
+{2V \over 3}  n_i n_{i+2} 
+{V \over 2}  n_i n_{i+3} \right ) 
\end{equation}
For a quarter-filled system, the exponent $\alpha$ of this model has already
been calculated with standard techniques\cite{capponi}, and it reaches the value
$1$ for $V/t=5$. Using Lanczos technique, we have calculated the dependence of 
$\Delta E$ on $t_\perp$ for different values of $V/t$ and different sizes up 
to 36 sites ($N=18$). Typical results are shown in Fig. 3.
The effects of intrachain repulsion are rather dramatic. The first oscillation
quickly becomes the dominant one, and the first value of $t_\perp$ where 
$\Delta E$ vanishes increases significantly: The curve is already strongly
affected for $V/t=2.5$ (Fig. 3) with respect to the non-interacting case 
(Fig. 2a). But more importantly, the oscillations disappear altogether
for $V/t=5$, i.e. $\alpha =1$. This critical value turns out to be independent 
of the size. The origin of the oscillations being that particles go from the
antibonding band to the bonding band, this means that these concepts have lost
their meaning when $\alpha$ is big enough. In other words, $t_\perp$ is no
longer able to produce two separate bands in the low energy spectrum of the
system. 

Our discussion of the dependence of $\Delta E$ on $N$ for interacting
systems is limited by the maximum size we can handle with Lanczos,
namely 36 sites ($N=18$) at quarter-filling. For $t_\perp=0.1$, we are limited
to the first linear section of Fig. 2b where the slope is equal to $2t_\perp$ in
the non-interacting case. The slope is considerably reduced by interactions and
{\it changes sign} between $V/t=5$ and $V/t=6$, which means that $\Delta E$ 
{\it decreases} with
$N$ (see Fig. 4a). Given the absence of oscillations 
in $\Delta E$ as a
function of $t_\perp$ for these values of the interaction, it is natural to 
assume that $\Delta E$ does
not change sign as a function of $N$ either. 
This leads us to the conclusion that 
$\Delta E$ goes to 0 in the thermodynamic limit when $\alpha \ge 1$\cite{note2}.
Significant effects are already present for $\alpha <1$, however. Let us
consider for instance $t_\perp=0.4$, for which $\Delta E$ changes sign between
$N=10$ and $N=14$ in the non-interacting case (see Fig. 4b). 
For the sizes we can
reach, $\Delta E$ already does not change sign for $V/t=2$, and the curve 
is very
flat for $V/t$ as small as 3 ($\alpha=0.46$). It is in fact possible that 
even in
that case $\Delta E$ goes to 0 with damped oscillations. Obtaining numerical
results on larger systems by other methods, e.g. Densitiy Matrix 
Renormalization
Group, would be very useful to check this point. Let us note that oscillations
would also be present for a regular ladder. However the periodicity remains
equal to $\phi_0$ even if $t_\perp$ is switched off. So the Moebius geometry is
essential to observe the effects of interactions.

\begin{figure}[hp]
\centerline{\psfig{figure=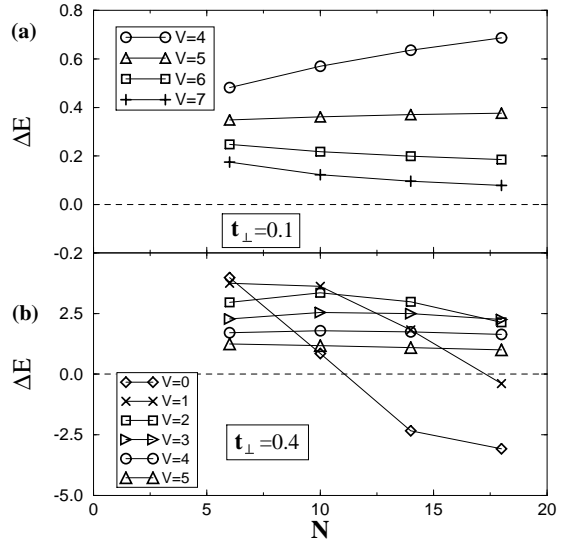,width=8.0cm,angle=0}}
\vspace{0.5cm}
\caption{$\Delta E$ as a function of the number of rungs $N$ for different
values of $V$ and two values of $t_\perp$: 
a) $t_\perp=0.1$; b) $t_\perp=0.4$. 
All energies are in units of $t$.}
\label{fig4}
\vspace{0.5cm}
\end{figure}

To make contact with current theories of the effect of interactions on
interchain hopping, let us first note that the present results strongly 
suggest that
the system has lost any memory of the splitting between bonding and antibonding
bands when $\alpha =1$. This is reminiscent of the RG
result that $t_\perp$ is an irrelevant perturbation when $\alpha>1$, but it is
much stronger: RG arguments are limited to infinitesimal values of
$t_\perp$, while our results show that {\it large} values of $t_\perp$ are still
unable to produce a difference between bonding and antibonding states if
$\alpha>1$. In fact, we believe that the results of Fig. 3 are the first direct
numerical evidence of a strong effect of interactions 
on hopping between chains
because they exhibit a qualitative change without having to go to the
thermodynamic limit. Besides, our results concerning the dependence of $\Delta
E$ on $N$ show very dramatic effects for relatively small interactions. 
These conclusions agree qualitatively with those of Refs.\cite{clarke} and
\cite{proceedings}, and they
are consistent with previous numerical studies of the spectral 
functions\cite{poilblanc,capponi}.

\begin{figure}[hp]
\centerline{\psfig{figure=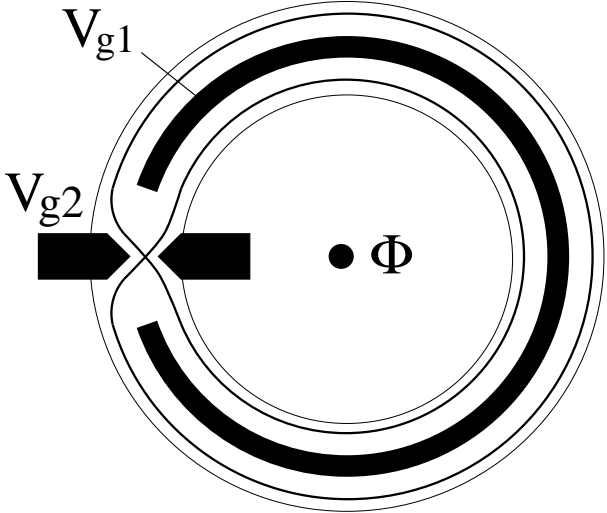,width=8.0cm,angle=0}}
\vspace{0.5cm}
\caption{Possible realization of a Moebius ring using a wide quantum 
wire.  At low densities, electrons will form Luttinger liquids along
the edges of the wire.  Interedge tunneling is controlled by
gate g1, while the crossover between the two edges is controlled by
gate g2.}
\label{fig5}
\end{figure}

TThe main advantage 
of the present approach is that the quantities we calculate
are not spectral functions, whose low-energy behaviour is very difficult to
measure with today's resolution, but thermodynamic quantities that should be 
accessible in experiments on mesoscopic systems.  A  Moebius ring with
controllable interedge tunneling could in principle be fabricated in a
GaAs heterostructure:  
At sufficiently low carrier
densities, electrons in a wide quantum wire will form Luttinger liquids
along the two edges, and 
interedge tunneling could be controlled via an external gate g1 (see Fig.\
5).
The crossover between the two edges could be
provided by a ballistic cross, controlled by gate g2, as illustrated in
Fig.\ 5. 
The persistent current in such a structure, which
is related to the ground state energy by $I(\Phi)=-(e/\hbar)\partial E/
\partial \Phi$, could be determined experimentally via magnetization
measurements \cite{currentexps}.  $I(\Phi)$ can be expressed
as a Fourier series, $I(\Phi)=\sum_{n=1}^{\infty} I_n \sin(n\Phi)$, and
such experiments typically measure the 
first two harmonics, $I_1$ and $I_2$.  The
first harmonic is in fact proportional to $\Delta E$, 
\begin{equation}
I_1 = (-1)^{\frac{n-1}{2}} \frac{e\Delta E}{2\hbar N} + {\cal O}(I_3),
\label{harmonic}
\end{equation}
provided the higher odd
harmonics can be neglected (which is always the case experimentally).
>From the above discussion of $\Delta E$, it is clear that $I_1$ should
be a highly sensitive function of $t_{\perp}$ and of the carrier density
in the coherent tunneling regime, while $I_1 = 0$ in the absence
of coherent interedge tunneling.  Experiments on suitably fabricated
mesoscopic systems, in which the predicted finite-size effects could
be directly measured, are thus good candidates to test theories of the
effect of interactions on interchain tunneling.

In conclusion, 
we have shown that interactions between electrons have dramatic
effects on their ability to hop coherently from one chain to the other by
studying the flux 
dependence of the ground-state energy -- or equivalently of 
the
persistent currents -- in a Moebius ladder. The absence of oscillations as a
function of $t_\perp$ for strong enough interactions is, we believe, the best
evidence of an interaction-induced destruction of interchain coherence
obtained so far with numerical simulations. Further
work along these lines, either numerically by studying larger systems, or
experimentally by measuring persistent currents in appropriate mesoscopic
devices, should be a promising area for future research.

We acknowledge useful discussions with D. Poilblanc. We thank IDRIS (Orsay) for 
allocation of CPU time on the C94 and C98 CRAY supercomputers.

\end{document}